\documentclass{article}

\usepackage{amsmath}
\usepackage{amssymb}
\usepackage{graphicx}
\usepackage{pdflscape}

\def\be{\begin{eqnarray}}
\def\ee{\end{eqnarray}}

\def\l[{\phantom.[}

\hoffset= - 1.0in         

\title{{On possible existence of HOMFLY polynomials for virtual knots} \vspace{.2cm}}
\author{{Alexei Morozov}\thanks{{\small
\textit{ITEP, Moscow, Russia}} and \textit{National Research Nuclear University MEPhI}; morozov@itep.ru},
{Andrey Morozov}\thanks{{\small
\textit{Moscow State University} and \textit{ITEP, Moscow, Russia} and \textit{Laboratory of Quantum Topology,
Chelyabinsk State University, Chelyabinsk, Russia} and \textit{National Research Nuclear University MEPhI}};
Andrey.Morozov@itep.ru}\,
and {Anton Morozov}\thanks{{\small \textit{
Moscow State University} and \textit{ITEP, Moscow, Russia}};
Anton.Morozov@itep.ru}}
\date{ }

\begin{document}

\maketitle

\vspace{-5cm}

\hfill ITEP/TH-23/14\\

\vspace{3.5cm}

\begin{abstract}
Virtual knots are associated with knot diagrams, which are not
obligatory planar.
The recently suggested generalization from $N=2$ to
arbitrary $N$ of the Kauffman-Khovanov calculus
of cycles in resolved diagrams can be straightforwardly applied
to non-planar case.
In simple examples we demonstrate that this construction preserves
topological invariance -- thus implying the existence of
HOMFLY extension of cabled Jones polynomials for
virtual knots and links.
\end{abstract}

\section{Introduction}

The main purpose of quantum field theory is evaluation of various
correlation functions in various models and understanding of their properties.
Especially interesting are non-perturbative (exact) results,
which exhibit a number of features, hidden in most perturbative expansions --
like dualities and integrability \cite{UFN2}.
The study in this direction is difficult because of the shortage of
examples, where reliable calculations can be performed:
they are currently restricted to models with high supersymmetry,
and to the closely connected conformal and Chern-Simons theories.
Any extension of knowledge in these fields is therefore very important,
any new family of calculable correlation functions is still very valuable.
In this letter we advocate the existence of new class of such quantities
in Chern-Simons theory -- these are HOMFLY polynomials for {\it virtual}
knots, a possible generalization of known theory in the direction of
non-simply-connected target spaces, where obstacles exist against the
use of the previously-developed technical tools, and thus essentially new
insights are expected to emerge.

\begin{picture}(100,110)(-120,-10)
\includegraphics[height=3cm,width=0.38\textwidth]{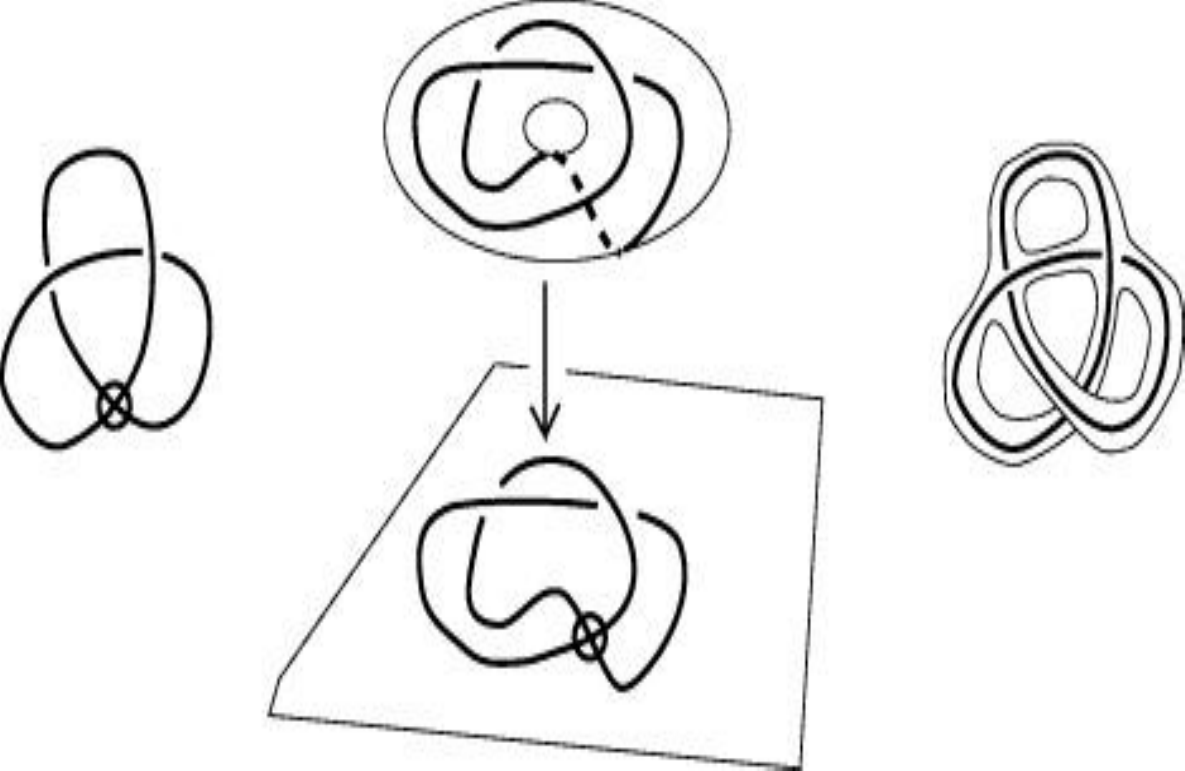}
\end{picture}


Kauffman's virtual links  and knots \cite{virknots}-\cite{tables}
are equivalence classes of link diagrams,
drawn on Riemann surfaces of arbitrary genus -- or, what is the same,
by non-obligatory-planar 4-valent graphs
(the picture shows the virtual trefoil $2.1$).
This means that in addition to black and white vertices, represented
by quantum ${\cal R}$-matrix and its inverse in
Reshetikhin-Turaev (RT) formalism \cite{RT}-\cite{AnoMcabling},
there are additional "sterile" crossings, marked by circles.
Despite their seeming simplicity, such crossings do not preserve
the quantum group representations -- and this makes application of RT
approach somewhat difficult.
What can be used, is alternative Kauffman's formalism \cite{KaufR},
based on the calculus of cycles in resolved diagrams,
closely related to Khovanov categorification approach \cite{Kho}.
However, in its standard form  \cite{BN,DM12} it is applicable only to
the case of $N=2$, i.e. the corresponding invariants (Jones polynomials) depend only on
parameter $q$ and a single representation label $r$, rather than
generic Young diagrams.
Still, on the base of Chern-Simons theory \cite{CS} one can expect
for virtual links and knots the existence of
knot polynomials \cite{knotpols} in arbitrary representation of arbitrary Lie algebra
-- in particular, of generic colored HOMFLY polynomials, in arbitrary
representation $R$ of arbitrary $SL(N)$ (with $N$-dependence absorbed into a variable
$A=q^N$).
In the absence of RT formalism these can be constructed with the help of
generalization of cycle calculus to $N>2$, suggested in \cite{DM3}.
However, since this approach is only in its very beginning,
rigorous presentation is not yet available.
The purpose of this letter  is to look through simple examples and demonstrate that
the polynomials, constructed in this way, are indeed topological invariants.
This provides a strong evidence that generalized-Reidemeister-invariant
HOMFLY polynomials $H^{\cal L}_{r}(q,A)$,
not only Jones $J^{\cal L}_r(q) = H^{\cal L}_{r}(q,q^2)$
can indeed exist for non-planar link/knot diagrams ${\cal L}$.
As to the possibility to extend {\it cabled} polynomials to arbitrary {\it colored} ones,
i.e. promote the number of wires in the cable $r$
to arbitrary representation (Young diagram) $R$, it remains obscure.

\section{Brief summary of  \cite{DM3}}

A link diagram ${\cal L}=\Gamma_c$ is an oriented graph $\Gamma$
(not planar, if the link is virtual) with 4-valent vertices
of two colors: black and white.

In RT approach one puts the quantum ${\cal R}$-matrix and its inverse at each black
and white vertex respectively and contract indices -- with additional insertion
of the grading factor $q^\rho$ at the upper turning points.

In Kauffman-Khovanov approach \cite{KaufR}-\cite{DM12} one proceeds differently.
We give a description directly in the version of  \cite{DM3},
which is relevant for our further consideration.

\subsection{Construction of HOMFLY polynomial from the hypercube of resolutions}

\noindent

1) Coloring of the graph $\Gamma$ is temporarily ignored -- till the step 7.

2) Instead each vertex is substituted by one of two possible {\it resolutions}.
With particular choice of the resolution one associates one of the $2^{n_\bullet+n_\circ}$
vertices of a hypercube ${\cal H}_\Gamma$.
The edge of the hypercube labels a switch of a resolution at one particular
vertex of $\Gamma$.

3) One of the two resolutions is trivial:
$\nearrow\!\!\!\!\!\!\nwarrow \ \longrightarrow \ \uparrow\uparrow$.
The hypercube has a distinguished ("Seifert") vertex $v_S$, where all the resolutions are trivial.

4) To begin with, we consider an auxiliary ("primary") hypercube, where the second
resolution is just $\nearrow\!\!\!\!\!\!\nwarrow \ \ \longrightarrow \ \
\nearrow\!\!\!\!\!\!\nwarrow\!\!\!\!\!\!\!\circ  $,
where a circle implies that the lines simply go through
(so that the graph gets additional non-planarity)
-- we call it {\it sterile} crossing.
At each vertex $v$ of this primary hypercube the graph $\Gamma$ is resolved into
a collection of $n_v$ cycles (perhaps, sterilely intersecting).
We call hypercube with these numbers $n_v$ at vertices $v$
the primary cycle hypercube (or diagram).

5) In the true hypercube the
non-trivial resolution is more sophisticated: it is a {\it difference} of two,
$\nearrow\!\!\!\!\!\!\nwarrow \ \ \longrightarrow \ \ \uparrow\uparrow \ -\  \nearrow\!\!\!\!\!\!\nwarrow
\!\!\!\!\!\!\!\circ  $.
Accordingly. more sophisticated is the number at the vertex $v$.
Namely, one should consider a sub-hupercube $C_{v,v_s}\subset{\cal H}_\Gamma$,
connecting $v$ with the Seifert vertex $v_S$
and take an alternated sum of $n_{v'}$ over all its vertices:
\be
D_v = \sum_{v'\in C_{v,v_S}} (-)^{|v'-v_S|} N^{n_{v'}}
\ee
where $|v'-v_S|$ is the distance (number of edges) between the vertex $v'$ and $v_S$
while $N$ is the extra parameter, interpreted as labeling of the $SL(N)$ algebra.
In Khovanov's categorification method $N_v$ is interpreted as dimensions of some
vector space -- in the context of \cite{DM3} it is rather a factor-space,
moreover, for virtual knots and links $N_v$ does not need to be positive.

6) The number $D_v$ should be "quantized" -- interpreted as "dimension" of a $q$-graded factor-space.
This is a subtle point: for ordinary knots and links the quantization receipt is actually
provided by ${\cal R}$-matrix calculus \cite{AnoM3}, but for virtual knots such technique
is not immediately available -- in the present paper we use non-rigorous mnemonic
quantization rules, like in \cite{DM3}.

7) Finally, to construct HOMFLY polynomial for original link diagram ${\cal L}=\Gamma_c$,
we associate original coloring $c$ of $\Gamma$
with particular ("initial") vertex $v_{\cal L}$ of the hypercube
(original black is associated with the trivial, while white -- with non-trivial resolution).
Then
\be
H^{\cal L}_{_\Box} = (-)^{n_\circ} q^{(N-1)n_\bullet-Nn_\circ}
\sum_{v \in {\cal H}_\Gamma} (-q)^{|v-v_{\cal L}|} D_v
\label{HOML}
\ee
where $n_\bullet$ and $n_\circ$ are the numbers of black and white vertices.
For totally-black coloring $v_{\cal L}=v_S$ and $q=1$ the alternated sum is just
$(-)^{n_\bullet+n_\circ} D_{\bar S}$ -- the classical dimension at the totally white
("anti-Seifert") vertex $v_{\bar S}$.

8) Cabled HOMFLY polynomials $H^{\cal L}_r$ are the fundamental HOMFLY for the the $r$-wire
cable, i.e. an $r$-component link (with wires additionally intertwined to make all the
pair linking numbers vanishing, see \cite{tables}).
For $N=2$ (Jones) this makes the set of cabled polynomials as big as that of the
colored ones, however, this is not true for $N>2$, when the number Young diagrams
of the size $r$ with $N-1$ rows exceeds $r$.
To define a richer family of {\it colored} HOMFLY polynomials one needs additional projectors,
like in \cite{AnoMcabling},
-- which are not yet available because of the lack of representation-respecting formalism.

9) In Khovanov-Rozansky theory \cite{Kho,KhR} eq.(\ref{HOML}) is further interpreted as Euler
characteristic of a certain complex, constructed with the help of cut-and-join
morphisms, acting along the edges of the hypercube.
Its Poincare polynomial is called Khovanov-Rozhansky polynomial,
and its stabilization at large enough $N$ is known as {\it superpolynomial}
\cite{GSV,DGR}.
A separate story is the proof of topological invariance of these quantities --
in the approach of \cite{DM3} is still remains to be found,
together with precise definition of cut-and-join morphisms.

\bigskip

Our convention for quantum numbers is
\be
\l[N] = \frac{q^N-q^{-N}}{q-q^{-1}}
\ee
To avoid possible confusion,
we emphasize that quantization of $D_v$ is more
involved than the substitution $D_v \longrightarrow [D_v]$.

\subsection{Example: ordinary (non-virtual) unknot}

\begin{picture}(100,80)(-40,-40)
\put(-20,-20){\line(1,1){40}}
\put(-20,20){\line(1,-1){40}}
\put(0,0){\circle*{5}}
\qbezier(20,20)(40,30)(40,0)
\qbezier(20,-20)(40,-30)(40,0)
\qbezier(-20,20)(-40,30)(-40,0)
\qbezier(-20,-20)(-40,-30)(-40,0)
\put(-10,-10){\vector(1,1){2}}
\put(10,-10){\vector(-1,1){2}}
\put(60,0){\vector(1,0){20}}
\put(120,-20){\line(1,1){15}}
\put(145,5){\line(1,1){15}}
\put(120,20){\line(1,-1){15}}
\put(145,-5){\line(1,-1){15}}
\qbezier(135,-5)(140,0)(135,5)
\qbezier(145,-5)(140,0)(145,5)
\qbezier(160,20)(180,30)(180,0)
\qbezier(160,-20)(180,-30)(180,0)
\qbezier(120,20)(100,30)(100,0)
\qbezier(120,-20)(100,-30)(100,0)
\put(130,-10){\vector(1,1){2}}
\put(150,-10){\vector(-1,1){2}}
\put(195,0){\mbox{$-\ q $}}
\qbezier(225,-30)(215,0)(225,30)
\put(250,-20){\line(1,1){15}}
\put(275,5){\line(1,1){15}}
\put(250,20){\line(1,-1){15}}
\put(275,-5){\line(1,-1){15}}
\qbezier(265,-5)(270,0)(265,5)
\qbezier(275,-5)(270,0)(275,5)
\qbezier(290,20)(310,30)(310,0)
\qbezier(290,-20)(310,-30)(310,0)
\qbezier(250,20)(230,30)(230,0)
\qbezier(250,-20)(230,-30)(230,0)
\put(260,-10){\vector(1,1){2}}
\put(280,-10){\vector(-1,1){2}}
\put(325,0){\mbox{$-$}}
\put(370,-20){\line(1,1){40}}
\put(370,20){\line(1,-1){40}}
\qbezier(410,20)(430,30)(430,0)
\qbezier(410,-20)(430,-30)(430,0)
\qbezier(370,20)(350,30)(350,0)
\qbezier(370,-20)(350,-30)(350,0)
\put(380,-10){\vector(1,1){2}}
\put(400,-10){\vector(-1,1){2}}
\qbezier(435,-30)(445,0)(435,30)
\end{picture}

\noindent

Primary cycle diagram (primary segment):   $2\longrightarrow 1$.

Hypercube with classical dimensions: $N^2 \longrightarrow N^2-N$.

Its quantization:  $[N]^2 \longrightarrow [N][N-1]$.

Unreduced HOMFLY:  $q^{N-1}\Big([N]^2-q[N][N-1]\Big) = q^{N-1}[N]\Big([N]-q[N-1]\Big) = [N]$.

Another choice of initial vertex: $-q^{-N}\Big([N][N-1]-q[N]^2\Big) =
q^{1-N}[N]\Big([N]-\frac{1}{q}[N-1]\Big) =[N]$.

Normalized HOMFLY: $1$.

\subsection{Virtual unknot}

\begin{picture}(100,80)(-150,-40)
\put(-20,-20){\line(1,1){40}}
\put(-20,20){\line(1,-1){40}}
\put(0,0){\circle{10}}
\qbezier(20,20)(40,30)(40,0)
\qbezier(20,-20)(40,-30)(40,0)
\qbezier(-20,20)(-40,30)(-40,0)
\qbezier(-20,-20)(-40,-30)(-40,0)
\put(-10,-10){\vector(1,1){2}}
\put(10,-10){\vector(-1,1){2}}
\end{picture}

Hypercube consists of a single vertex and unreduced HOMFLY polynomial is $[N]$,
while reduced is just $1$.
Thus we define these polynomials for the virtual unknot to be the same as for the
ordinary unknot -- like in \cite{tables}.

\section{Example of topological invariance: virtual trefoil}

In this section we provide the first illustration
that HOMFLY polynomials {\it a la} \cite{DM3} are indeed
topological invariants for virtual knots.
Namely we consider virtual trefoil in two different
--$2$-strand and $3$-strand -- realizations,
calculate their HOMFLY polynomials
and see that they coincide.

\subsection{Virtual trefoil (2-strand version): $2.1$ in the notation of \cite{tables}}

\begin{picture}(100,100)(-150,-50)
\qbezier(0,0)(20,30)(40,0)
\qbezier(0,0)(20,-30)(40,0)
\qbezier(40,0)(60,30)(80,0)
\qbezier(40,0)(60,-30)(80,0)
%
\qbezier(0,0)(-11,15)(5,25)
\qbezier(80,0)(91,15)(75,25)
\qbezier(5,25)(40,55)(75,25)
\qbezier(0,0)(-11,-15)(5,-25)
\qbezier(80,0)(91,-15)(75,-25)
\qbezier(5,-25)(40,-55)(75,-25)
\put(41,40){\vector(-1,0){2}}
\put(41,-40){\vector(-1,0){2}}
\put(40,0){\circle{10}}
\put(0,0){\circle*{5}}
\put(80,0){\circle*{5}}
\end{picture}

{\bf Primary cycle diagram} (primary square):
$$
\begin{array}{ccccc}
&&2&& \\
&\nearrow&&\searrow \\
1&&&&1 \\
&\searrow &&  \nearrow \\
&& 2 &&
\end{array}
$$

{\bf Classical "dimensions":}
$$
\begin{array}{ccccc}
&&N-N^2&& \\
&\nearrow&&\searrow \\
N&&&&2N-2N^2 \\
&\searrow &&  \nearrow \\
&& N-N^2 &&
\end{array}
$$
Note\ that
$$
\!\!\!\!\!\!\!\!\!\!\!\!\!\!\!\!\!\!\!\!\!\!\!\!\!
\boxed{ {\rm  for\ virtual\ knots\ "dimensions"\ can\ be\ negative}}
$$

{\bf Quantization:}
$$
\begin{array}{ccccc}
&&-[N][N-1]&& \\
&\nearrow&&\searrow \\
\l[N]&&&&-[2]N][N-1] \\
&\searrow &&  \nearrow \\
&& -[N][N-1] &&
\end{array}
$$

{\bf Unreduced HOMFLY:}
\be
H^{2.1}_{_\Box}(q)=q^{2(N-1)}\Big([N] -2q(-[N][N-1])+q^2(-[2][N][N-1]\Big)
=[N]\Big(-q^{3N-1}+q^{2N-2}+q^{N+1})\Big)
\label{H2.1}
\ee
In particular, for $N=2$ we get the answer from \cite{virknots,tables}:
\be
J^{2.1}_{_\Box}= [2]\Big(-q^5+q^3+q^{2}\Big)
\label{J2.1}
\ee

{\bf The opposite initial vertex:}
$$
(-q^{-N})^2\Big(-[2][N][N-1]-2q(-[N][N-1])+q^2[N]\Big)
= [N]\Big(-q^{1-3N}+q^{2-2N}+q^{-1-N}\Big)
\stackrel{(\ref{H2.1})}{=} H^{2.1}_{_\Box}(q^{-1})
$$

{\bf Alternative initial vertex} (unknot expected):
\be
-q^{-N}\cdot q^{N-1}\Big(-[N][N-1]-q([N]-[2][N][N-1])+q^2(-[N][N-1])\Big)
=-q^{-1}[N]\cdot(-q)=[N]
\ee

\subsection{Three comments}

At least three properties of the answer (\ref{H2.1}) deserve attention.

1) Polynomials for virtual knots contain odd powers of $q$.

2) The quantities (quantum "dimensions") at the hupercube vertices
can be negative.
This is easily conceivable in the approach of \cite{DM3}, where non-trivial
resolution is a {\it difference}.
However, this seems impossible in the standard Kauffman's approach \cite{KaufR}
at $N=2$, where dimensions are just powers of $[2]$ (see \cite{BN,DM12} for details).
In order to obtain the {\it right} answer (\ref{J2.1}) -- which follows
immediately from (\ref{H2.1}) -- from the standard approach with the resolutions
$\uparrow\uparrow$ and  $\stackrel{\cup}{\cap}$,
an artificially-looking substitution
$[2]\longrightarrow -[2]$ (or, what can be equivalent, $q\longrightarrow -q$)
had to be made "by hands" in the original paper -- the first one in ref.\cite{virknots}.
Examination of the general-$N$ case in the framework of \cite{DM3}
provides a natural explanation for this trick.

3) The answer (\ref{H2.1}) contains three items, separated by factors $\sim q^N$
rather than $\sim q^{2N}$.
This makes it impossible to decompose this formula into a linear combination
of $[N+1]$ and $[N-1]$ and thus to interpret it as a combination of quantum
(graded) dimensions $\frac{[N][N\pm 1]}{[2]}$ of symmetric and antisymmetric
representations of $SL(N)$.
This reflects the problems with naive application of RT approach to virtual knots.

\subsection{Virtual trefoil (3-strand version)}

\noindent

{\bf Primary cycle diagram:}
$$
\begin{array}{ccccccc}
&&1&\rightarrow& 2&& \\
&\nearrow&&\nearrow\!\!\!\!\!\!\searrow&&\searrow \\
2&\rightarrow&3&&2&\rightarrow & 1 \\
&\searrow &&\nearrow\!\!\!\!\!\!\searrow&&  \nearrow \\
&& 1 &\rightarrow& 2
\end{array}
$$

{\bf Classical "dimensions":}
$$
\begin{array}{ccccccc}
&&N^2-N&\rightarrow& 2N^2-N-N^3&& \\
&\nearrow&&\nearrow\!\!\!\!\!\!\searrow&&\searrow \\
N^2&\rightarrow&N^2-N^3&&2N^2-2N&\rightarrow & 4N^2-3N-N^3=-\boxed{N(N-1)(N-3)} \\
&\searrow &&\nearrow\!\!\!\!\!\!\searrow&&  \nearrow \\
&& N^2-N &\rightarrow& 2N^2-N-N^3
\end{array}
$$

{\bf Quantization} is non-trivial only for the boxed item:
as demonstrated in \cite{DM3}, there should be no "gaps" in the products,
i.e. $N-3$ should rather be substituted by some linear combination of
$N-1$, $N-2$ and $1$.  For the time-being we denote the quantization of $N-3$ by $D$:
$$
\begin{array}{ccccccc}
&&[N][N-1]&\rightarrow& -[N][N-1]^2&& \\
&\nearrow&&\nearrow\!\!\!\!\!\!\searrow&&\searrow \\
\l[N]^2&\rightarrow&\underline{\underline{-[N]^2[N-1]}}&&[2][N][N-1]&\rightarrow & -[N][N-1]D \\
&\searrow &&\nearrow\!\!\!\!\!\!\searrow&&  \nearrow \\
&& \underline{[N][N-1]} &\rightarrow& -[N][N-1]^2
\end{array}
$$

As we shall now see, one and
the same $D=[N-2]-1$ will match {\it both} the unknot and the trefoil.
Thus one can say. that $D$ is defined from the requirement that the unknot
is properly described -- while the answer for the virtual virtual trefoil
in the 3-strand representation is {\it deduced}.
Also predicted will be HOMFLY polynomials for the virtual figure-eight knot,
which emerges if another hypercube vertex is taken for initial one.

{\bf Unreduced HOMFLY:}
$$
q^{3(N-1)}\left\{[N]^2-q\Big(2[N][N-1]-[N]^2[N-1]\Big)
+q^2\Big([2][N][N-1]-2[N][N-1]^2\Big)-q^3\Big(-[N][N-1]D\Big)\right\}=
$$
\vspace{-0.7cm}
\be
= [N]\Big(-q^{3N-1}+q^{2N-2}+q^{N+1}\Big)
\ \stackrel{(\ref{H2.1})}{=}\
H^{2.1}_{_\Box}(q)
\ee

$$
\boxed{
{\rm
Thus\ HOMFLY\ polynomials\ in\ two\ different\ realizations\ of\ the\
same\ virtual\ knot\ are\ indeed\ the\ same.
}}
$$

\bigskip

{\bf Alternative initial vertex (underlined) -- the unknot:}
$$
-q^{-N}q^{2(N-1)}\Big\{[N-1]-q\Big([N]-[N-1]^2+[2][N-1]\Big)
+ q^2\Big(-D+[N-1]-[N][N-1]\Big)
- q^3\Big(-[N-1]^2\Big)\Big\} = 1
$$

{\bf Another alternative initial vertex (double-underlined) -- the virtual figure eight
($3.2$ of \cite{tables}):}
$$
-q^{-N}q^{2(N-1)}\Big\{-[N]^2[N-1]-q\Big([N]^2-2[N][N-1]^2\Big)+q^2\Big(2[N][N-1]-[N][N-1]D\Big)
-q^3[2][N][N-1]\Big\}
$$
i.e.
\be
H_{_\Box}^{3.2}
= [N]\Big(q^{2N}-q^{N-1}-q^2+1+q^{1-N}\Big)
\ee
For $N=2$ this turns into the known answer from \cite{tables}:
\be
J_{_\Box}^{3.2} = [2]\Big(q^4-q^2-q+1+q^{-1}\Big)
\ee

\newpage

\begin{landscape}

\section{The list of HOMFLY polynomials for the simplest virtual knots}

For the 2- and 3-intersection virtual knots from \cite{tables}
we obtain in this way the following reduced polynomials:

\bigskip

$$
\begin{array}{|c|c|c|c|c|c|}
\hline
&&&&&\\
\text{Knot} & \text{Diagram} & \text{Jones} &  \text{HOMFLY} & \text{Primary cycle diagram} & \text{Normalized quantum ``dimensions''} \\
&&&&&\\
\hline
&&&&& \\
2.1 & \begin{picture}(10,10)(10,10)
\includegraphics[width=1cm,height=1cm]{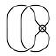}
\end{picture} & -q^5+q^3+q^2   &   -\cfrac{A^3}{q}+\cfrac{A^2}{q^2}+Aq &
\begin{array}{ccccc}
&&2&& \\
&\nearrow&&\searrow& \\
\boxed{1}&&&&1 \\
&\searrow &&  \nearrow & \\
&&2&&
\end{array}
&
\footnotesize{
\begin{array}{ccccc}
&&-[N-1] && \\
&\nearrow&&\searrow & \\
\boxed{1}&&&&-[2][N-1] \\
&\searrow &&  \nearrow &\\
&&-[N-1] &&
\end{array}
}
\\
&&&&& \\
\hline
&&&&& \\
3.1 &
\begin{picture}(10,10)(10,10)
\includegraphics[width=1cm,height=1cm]{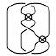}
\end{picture}
& 1 & 1 &
\begin{array}{ccccccc}
&&\boxed{1}&\rightarrow& 2&& \\
&\nearrow&&\nearrow\!\!\!\!\!\!\searrow&&\searrow \\
2&\rightarrow&3&&2&\rightarrow & 1 \\
&\searrow &&\nearrow\!\!\!\!\!\!\searrow&&  \nearrow \\
&& 1 &\rightarrow& 2
\end{array}
&
\setlength{\arraycolsep}{0pt}
\footnotesize{
\begin{array}{ccccccc}
&&\boxed{[N-1]}&\rightarrow& -[N-1]^2&& \\
&\nearrow&&\nearrow\!\!\!\!\!\!\searrow&&\searrow \\
\l[N]&\rightarrow&-[N][N-1]&&[2][N-1]&\rightarrow & -[N-1]\Big([N-2]-1\Big) \\
&\searrow &&\nearrow\!\!\!\!\!\!\searrow&&  \nearrow \\
&& [N-1] &\rightarrow& -[N-1]^2
\end{array}
}
\setlength{\arraycolsep}{6pt}
\\
&&&&& \\
\hline
&&&&& \\
3.2 &
\begin{picture}(10,10)(10,10)
\includegraphics[width=1cm,height=1cm]{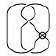}
\end{picture} & q^4-q^2-q+1+\cfrac{1}{q}   &  A^2-\cfrac{A}{q}-q^2+1+\cfrac{q}{A} &
\begin{array}{ccccccc}
&&1&\rightarrow& 2&& \\
&\nearrow&&\nearrow\!\!\!\!\!\!\searrow&&\searrow \\
2&\rightarrow&\boxed{3}&&2&\rightarrow & 1 \\
&\searrow &&\nearrow\!\!\!\!\!\!\searrow&&  \nearrow \\
&& 1 &\rightarrow& 2
\end{array}
&
\setlength{\arraycolsep}{0pt}
\footnotesize{
\begin{array}{ccccccc}
&&[N-1]&\rightarrow& -[N-1]^2&& \\
&\nearrow&&\nearrow\!\!\!\!\!\!\searrow&&\searrow \\
\l[N]&\rightarrow&\boxed{-[N][N-1]}&&[2][N-1]&\rightarrow & -[N-1]\Big([N-2]-1\Big) \\
&\searrow &&\nearrow\!\!\!\!\!\!\searrow&&  \nearrow \\
&& [N-1] &\rightarrow& -[N-1]^2
\end{array}
}
\setlength{\arraycolsep}{6pt}
\\
&&&&& \\
\hline
&&&&& \\
3.3 &
\begin{picture}(10,10)(10,10)
\includegraphics[width=1cm,height=1cm]{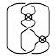}
\end{picture} & -q^5+q^3+q^2 & -\cfrac{A^3}{q}+\cfrac{A^2}{q^2}+Aq &
\begin{array}{ccccccc}
&&1&\rightarrow& 2&& \\
&\nearrow&&\nearrow\!\!\!\!\!\!\searrow&&\searrow \\
\boxed{2}&\rightarrow&3&&2&\rightarrow & 1 \\
&\searrow &&\nearrow\!\!\!\!\!\!\searrow&&  \nearrow \\
&& 1 &\rightarrow& 2
\end{array}
&
\setlength{\arraycolsep}{0pt}
\footnotesize{
\begin{array}{ccccccc}
&&[N-1]&\rightarrow& -[N-1]^2&& \\
&\nearrow&&\nearrow\!\!\!\!\!\!\searrow&&\searrow \\
\boxed{[N]}&\rightarrow&-[N][N-1]&&[2][N-1]&\rightarrow & -[N-1]\Big([N-2]-1\Big) \\
&\searrow &&\nearrow\!\!\!\!\!\!\searrow&&  \nearrow \\
&& [N-1] &\rightarrow& -[N-1]^2
\end{array}
}
\setlength{\arraycolsep}{6pt}
\\&&&&&\\
\hline
\end{array}
$$

$$
\begin{array}{|c|c|c|c|c|c|}
\hline
&&&&&\\
\text{Knot} & \text{Diagram} & \text{Jones} &  \text{HOMFLY} & \text{Primary cycle diagram} & \text{Normalized quantum ``dimensions''} \\
&&&&&\\
\hline
&&&&& \\
3.4 & \begin{picture}(10,10)(10,10)
\includegraphics[width=1cm,height=1cm]{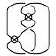}
\end{picture} & q^4-q^2-q+1+\cfrac{1}{q}  &   A^2-\cfrac{A}{q}-q^2+1+\cfrac{q}{A} &
\begin{array}{ccccccc}
&&1&\rightarrow& 2&& \\
&\nearrow&&\nearrow\!\!\!\!\!\!\searrow&&\searrow \\
2&\rightarrow&\boxed{3}&&2&\rightarrow & 1 \\
&\searrow &&\nearrow\!\!\!\!\!\!\searrow&&  \nearrow \\
&& 1 &\rightarrow& 2
\end{array}
&
\setlength{\arraycolsep}{0pt}
\footnotesize{
\begin{array}{ccccccc}
&&[N-1]&\rightarrow& -[N-1]^2&& \\
&\nearrow&&\nearrow\!\!\!\!\!\!\searrow&&\searrow \\
\l[N]&\rightarrow&\boxed{-[N][N-1]}&&[2][N-1]&\rightarrow & -[N-1]\Big([N-2]-1\Big) \\
&\searrow &&\nearrow\!\!\!\!\!\!\searrow&&  \nearrow \\
&& [N-1] &\rightarrow& -[N-1]^2
\end{array}
}
\setlength{\arraycolsep}{6pt}
\\
&&&&& \\
\hline
&&&&& \\
3.5 & \begin{picture}(10,10)(10,10)
\includegraphics[width=1cm,height=1cm]{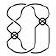}
\end{picture} & -q^8+q^6+q^2   &  -A^4+A^2\left(q^2+q^{-2}\right) &
\begin{array}{ccccccc}
&&1&\rightarrow& 2&& \\
&\nearrow&&\nearrow\!\!\!\!\!\!\searrow&&\searrow \\
\boxed{2}&\rightarrow&1&&2&\rightarrow & 1 \\
&\searrow &&\nearrow\!\!\!\!\!\!\searrow&&  \nearrow \\
&& 1 &\rightarrow& 2
\end{array}
&
\setlength{\arraycolsep}{0pt}
\footnotesize{
\begin{array}{ccccccc}
&&[N-1]&\rightarrow&[2][N-1]&& \\
&\nearrow&&\nearrow\!\!\!\!\!\!\searrow&&\searrow \\
\boxed{[N]}&\rightarrow&[N-1]&&[2][N-1]&\rightarrow &[2]^2[N-1] \\
&\searrow &&\nearrow\!\!\!\!\!\!\searrow&&  \nearrow \\
&&[N-1]&\rightarrow&[2][N-1]
\end{array}
}
\setlength{\arraycolsep}{6pt}
\\
&&&&& \\
\hline
&&&&& \\
3.6 & \begin{picture}(10,10)(10,10)
\includegraphics[width=1cm,height=1cm]{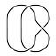}
\end{picture} & -q^8+q^6+q^2   &  -A^4+A^2\left(q^2+q^{-2}\right) &
\begin{array}{ccccccc}
&&1&\rightarrow& 2&& \\
&\nearrow&&\nearrow\!\!\!\!\!\!\searrow&&\searrow \\
\boxed{2}&\rightarrow&1&&2&\rightarrow & 1 \\
&\searrow &&\nearrow\!\!\!\!\!\!\searrow&&  \nearrow \\
&& 1 &\rightarrow& 2
\end{array}
&
\setlength{\arraycolsep}{0pt}
\footnotesize{
\begin{array}{ccccccc}
&&[N-1]&\rightarrow&[2][N-1]&& \\
&\nearrow&&\nearrow\!\!\!\!\!\!\searrow&&\searrow \\
\boxed{[N]}&\rightarrow&[N-1]&&[2][N-1]&\rightarrow &[2]^2[N-1] \\
&\searrow &&\nearrow\!\!\!\!\!\!\searrow&&  \nearrow \\
&&[N-1]&\rightarrow&[2][N-1]
\end{array}
}
\setlength{\arraycolsep}{6pt}
\\
&&&&& \\
\hline
&&&&& \\
3.7 & \begin{picture}(10,10)(10,10)
\includegraphics[width=1cm,height=1cm]{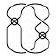}
\end{picture} & 1 &  1 &
\begin{array}{ccccccc}
&&1&\rightarrow& 2&& \\
&\nearrow&&\nearrow\!\!\!\!\!\!\searrow&&\searrow \\
2&\rightarrow&1&&2&\rightarrow & 1 \\
&\searrow &&\nearrow\!\!\!\!\!\!\searrow&&  \nearrow \\
&& \boxed{1} &\rightarrow& 2
\end{array}
&
\setlength{\arraycolsep}{0pt}
\footnotesize{
\begin{array}{ccccccc}
&&[N-1]&\rightarrow&[2][N-1]&& \\
&\nearrow&&\nearrow\!\!\!\!\!\!\searrow&&\searrow \\
\l [N]&\rightarrow&[N-1]&&[2][N-1]&\rightarrow &[2]^2[N-1] \\
&\searrow &&\nearrow\!\!\!\!\!\!\searrow&&  \nearrow \\
&&\boxed{[N-1]}&\rightarrow&[2][N-1]
\end{array}
}
\setlength{\arraycolsep}{6pt}
\\&&&&&\\
\hline
\end{array}
$$

\bigskip

\noindent
The list and the pictures are taken from \cite{tables},
Jones polynomials are obtained by putting $N=2$ in HOMFLY --
they coincide with those in \cite{tables}, up to the usual
change of variable $q \longrightarrow q^{\pm 1/2}$.
Quantization follows the general rules from \cite{DM3}.
$N$-dependence of HOMFLY is captured into  $A=q^N$.
The overall factor $[N]$ is omitted in hypercubes and knot
polynomials.
In the hypercubes the boxed items denote initial vertices.

\end{landscape}

\newpage


With a single exception,
in these examples the fundamental HOMFLY do \textit{not}  distinguish
virtual knots, which are not distinguished by Jones polynomials.
As one can see from \cite{tables}, one needs \textit{cabled} polynomials
to establish the difference -- this is similar to using
\textit{colored} HOMFLY to distinguish, say, mutant knots \cite{Mor}.
However in this latter example the non-symmetric representations
were needed.
Since cabled polynomials look like getting contributions from
\textit{all} representations of a given level, it remains a question,
what "constituent" of the cabling really matters in the virtual case.



Note that  coincidence of HOMFLY for $3.2$ and $3.4$ and for $3.5$ and $3.6$
follows from coincidence of their hypercubes and initial vertices,
while for $3.1$ and $3.7$ the hypercubes are different,
still, fundamental HOMFLY are the same (and coincide with that for the
unknot -- this is known for Jones since the original papers in \cite{virknots}
but remains true for arbitrary $N$, only cabled polynomials reveal the
difference). For $2.1$ and $3.7$ fundamental HOMFLY is already enough to
distinguish knots, which were not distinguished by Jones polynomial.

For $3.5$ and $3.7$ the hypercubes are just the same as for the
ordinary (non-virtual) knot $3.6$ (in particular all the dimensions
are positive). Thus it is not a surprise that HOMFLY and Jones in these
cases do not contain odd powers of $q$ or $A$.
However, this argument is not enough to explain the same
property in the case of $3.1$.

\section{Generic 2-strand case}

It is instructive to compare two {\it families} of knots: one ordinary and one virtual.
In the first case the $2$-strand braid consists of $2n+1$ black vertices.
In the second case one of them is substituted by a sterile crossing.

\begin{picture}(100,100)(0,-50)
\qbezier(0,0)(20,30)(40,0)
\qbezier(0,0)(20,-30)(40,0)
\qbezier(40,0)(60,30)(85,-5)
\qbezier(40,0)(60,-30)(85,5)
\put(95,0){\mbox{$\ldots$}}
\qbezier(115,-5)(140,30)(160,0)
\qbezier(115,5)(140,-30)(160,0)
\qbezier(160,0)(180,30)(200,0)
\qbezier(160,0)(180,-30)(200,0)
%
\qbezier(0,0)(-11,15)(5,25)
\qbezier(200,0)(211,15)(195,25)
\qbezier(5,25)(100,55)(195,25)
\qbezier(0,0)(-11,-15)(5,-25)
\qbezier(200,0)(211,-15)(195,-25)
\qbezier(5,-25)(100,-55)(195,-25)
\put(101,40){\vector(-1,0){2}}
\put(101,-40){\vector(-1,0){2}}
\put(0,0){\circle*{5}}
\put(40,0){\circle*{5}}
\put(80,0){\circle*{5}}
\put(120,0){\circle*{5}}
\put(160,0){\circle*{5}}
\put(200,0){\circle*{5}}
\end{picture}
\begin{picture}(100,100)(-150,-50)
\qbezier(0,0)(20,30)(40,0)
\qbezier(0,0)(20,-30)(40,0)
\qbezier(40,0)(60,30)(85,-5)
\qbezier(40,0)(60,-30)(85,5)
\put(95,0){\mbox{$\ldots$}}
\qbezier(115,-5)(140,30)(160,0)
\qbezier(115,5)(140,-30)(160,0)
\qbezier(160,0)(180,30)(200,0)
\qbezier(160,0)(180,-30)(200,0)
%
\qbezier(0,0)(-11,15)(5,25)
\qbezier(200,0)(211,15)(195,25)
\qbezier(5,25)(100,55)(195,25)
\qbezier(0,0)(-11,-15)(5,-25)
\qbezier(200,0)(211,-15)(195,-25)
\qbezier(5,-25)(100,-55)(195,-25)
\put(101,40){\vector(-1,0){2}}
\put(101,-40){\vector(-1,0){2}}
\put(0,0){\circle{10}}
\put(40,0){\circle*{5}}
\put(80,0){\circle*{5}}
\put(120,0){\circle*{5}}
\put(160,0){\circle*{5}}
\put(200,0){\circle*{5}}
\end{picture}

\noindent
In somewhat symbolical notation the primary hypercubes in these two cases are respectively

$$
2 \longrightarrow (2n)\cdot \underline{1} \longrightarrow \frac{(2n)(2n-1)}{2}\cdot \underline{2}
\longrightarrow \ \ldots\ \longrightarrow C^{2i}_{2n}\cdot \underline{2}
\longrightarrow C^{2i+1}_{2n}\cdot \underline{1} \longrightarrow \ \ldots \ \longrightarrow \underline{2}
$$
and
$$
1 \longrightarrow (2n)\cdot \underline{2} \longrightarrow \frac{(2n)(2n-1)}{2}\cdot \underline{1}
\longrightarrow \ \ldots\ \longrightarrow C^{2i}_{2n}\cdot \underline{1}
\longrightarrow C^{2i+1}_{2n}\cdot \underline{2} \longrightarrow \ \ldots \ \longrightarrow \underline{1}
$$

\noindent
i.e. where there was one cycle in one case there are two cycles in another case
and vice versa.
Concerning notation, underlined are the numbers of cycles ($n_v$), and coefficients
in front of them are the numbers of vertices with the same $n_v$.

However, these two configurations lead to rather different hypercubes:

$$
\underline{[N]^2} \longrightarrow (2n)\cdot \underline{[N][N-1]} \longrightarrow \frac{(2n)(2n-1)}{2}\cdot \underline{[2][N][N-1]}
\longrightarrow \ \ldots\
$$
$$
\ldots\ \longrightarrow C^{2i}_{2n}\cdot \underline{[2]^{2i-1}[N][N-1]}
\longrightarrow C^{2i+1}_{2n}\cdot \underline{[2]^i[N][N-1]} \longrightarrow \ \ldots \ \longrightarrow
\underline{[2]^{2n-1}[N][N-1]}
$$
and
$$
[N] \longrightarrow (2n)\cdot \underline{\big(-[N][N-1]\big)} \longrightarrow
\underline{\frac{(2n)(2n-1)}{2}}\cdot \underline{\big(-[2][N][N-1]\big)}
\longrightarrow \ \ldots\
$$
$$
\ldots \ \longrightarrow C^{2i}_{2n}\cdot \underline{\big(-[2]^{2i-1}[N][N-1]\big)}
\longrightarrow C^{2i+1}_{2n}\cdot \underline{\big(-[2]^i[N][N-1]\big)} \longrightarrow \ \ldots \ \longrightarrow
\underline{\big(-[2]^{2n-1}[N][N-1]\big)}
$$

\bigskip

\noindent
Thus different are the resulting unreduced HOMFLY polynomials:
$$
q^{(2n+1)(N-1)}[N]\left(\underline{[N]}\ +\
\sum_{i=1}^{2n+1}  C^i_{2n}(-q)^i \underline{[2]^{i-1}[N-1]}\right)
=  q^{(2n+1)(N-1)}\left([N]
\ +\ \Big(\big(1-q[2]\big)^{2n+1}-1\Big)\frac{ [N][N-1]}{[2]}\right) =
$$
\vspace{-0.4cm}
\be
= q^{(2n+1)N}\left\{q^{-2n-1}\, \frac{[N][N+1]}{[2]} - q^{2n+1}\, \frac{[N][N-1]}{[2]}\right\}
\ee
and
$$
q^{(2n+1)(N-1)}[N]\left(\underline{1}\ +\
\sum_{i=1}^{2n}  C^i_{2n}(-q)^i \Big(\underline{-[2]^{i-1}[N-1]}\Big)\right)
= q^{2n(N-1)}[N]\left(1
\ -\ \Big(\big(1-q[2]\big)^{2n}-1\Big)\frac{ [N-1]}{[2]}\right) =
$$
\vspace{-0.4cm}
\be
\boxed{
= q^{2nN}\left\{q^{-2n}\, \left(\frac{[N][N-1]}{[2]} + [N]\right) - q^{2n}\, \frac{[N][N-1]}{[2]}
\right\}^{\phantom{2^2}}_{\phantom{_2}}
}
\label{virt2str}
\ee
For $N=2$ and $n=1,2$ this reproduces the Jones polynomials for virtual knots $2.1$ and $4.100$
of \cite{tables}.

\bigskip

Comparing the two formulas, one can see that the $n$-dependence in both cases is nicely
described by the RT-inspired evolution method of \cite{DMMSS,evo},
with the ${\cal R}$-matrix eigenvalues $A/q=q^{N-1}$ and $-Aq=-q^{N+1}$ in
symmetric and antisymmetric representations.
Moreover, it looks like the additional crossing operator preserves
the structure of antisymmetric representation (the corresponding eigenvalue is,
of course, $-1$) -- at least the quantum (graded) dimension $\frac{[N][N-1]}{[2]}$
remains intact.
However, the structure of symmetric representation is destroyed:
the quantum dimension is changed from the usual $\frac{[N][N+1]}{[2]}$
to somewhat mysterious, still inspiring  $\frac{[N][N-1]}{[2]} + [N]$,
implying the special role of diagonal matrices.
A very interesting next question is what happens to the mixing (Racah)
matrices of \cite{knMMM12} for three and more strands.
The answer to this question can be crucial for existence of some
modified RT calculus for virtual knots.

%


\section{Conclusion}

In this paper we applied the technique of \cite{DM3} to virtual links and knots,
and presented evidence that this allows to lift the known Jones polynomials
to HOMFLY, depending on one extra parameter $A=q^N$ -- and these extended quantities
are also topological invariants.
Of course, this opens a new chapter in the study of virtual knots.
On the other hand, this sheds fresh light on the general theory of knot polynomials,
because generalization from ordinary to virtual knots breaks numerous
properties of the standard calculus:
representation theory and thus RT method are not applicable, at least straightforwardly,
polynomials break $q\leftrightarrow -q$ symmetry, "dimensions" of vector spaces
at hypercube vertices can be negative, thus causing certain problems in
the definition of Khovanov-Rozansky and super-polynomials.
Surprisingly or not, the approach of \cite{DM3} seems to survive in this shaky situation
and at the moment looks like the only viable possibility to define
rich enough knot polynomials for virtual knots.

In application to HOMFLY this formalism consists of two steps:
calculating the numbers of cycles for different resolutions of the link diagram
and then quantizing these numbers, by making a $q$-deformation or a $q$-grading,
depending on preferred language and interpretation.
For ordinary knots and links there is at least one way to make this quantization
rigorously and unambiguously -- with the help of ${\cal R}$-matrix calculus,
as described in \cite{AnoM3}.
However, representation-theory interpretation of "sterile" crossings is still
unavailable -- thus, when they are present,
this technical advance of \cite{AnoM3} is no longer applicable.
Still, getting more formulas like the unexpectedly inspiring (\ref{virt2str})
can hopefully allow
to understand, how RT approach can be modified to include
sterile crossings -- and thus to {\it derive} quantization rules from the first principles.
Of course in the absence of sterile crossings we obtain just the usual
HOMFLY polynomials for ordinary knots and links.

Another interesting question is the Chern-Simons theory description
of virtual knots.
Since such knots can be considered as embedded into non-simply-connected
$3d$ space, it seems that the Wilson-loop averages can depend on
{\it additional} free parameters:
monodromies around non-contractable cycles on underlying Riemann surface.
These parameters were ignored in \cite{virknots}, but there is also no room
for them in the framework of the present paper.
In particular in s.2.3 we demonstrated that our HOFMLY for the virtual unknot
is just the same as for the ordinary one, i.e. the possible dependence on
monodromies is indeed ignored.

Also open is the question about the possibility to define some analogue
of Khovanov-Rozansky \cite{KhR,DM3}
and super-polynomials \cite{GSV}-\cite{Dan} for virtual knots.

All this makes the situation both intriguing and promising.
The study of cabled HOMFLY polynomials for virtual links and knots
and their further generalizations
is clearly going to provide us with new and important insights.

\section*{Acknowledgements}

We are indebted to A.Mironov for attracting our attention to the
subject of virtual knots and to D.Bar-Natan for valuable comments
on the subject.
Our work is partly supported by grant NSh-1500.2014.2 (A.M. and And.M.), by RFBR grants 13-02-00478 (A.M.),
14-01-00547 (And.M.), 14-01-31395\_young\_a (And.M.), 12-01-00482 (Ant.M.), 14-02-31446\_young\_a (Ant.M.), by joint grants 13-02-91371-ST and 14-01-92691-Ind (A.M. and And.M.), by the Brazil National Counsel of Scientific and Technological Development (A.M.), by the Laboratory of Quantum Topology of Chelyabinsk State University (Russian Federation government grant 14.Z50.31.0020) (And.M.) and by the Dynasty Foundation (And.M.).

\end{document}